\documentclass[preprint,prd,aps,showpacs,showkeys,onecolumn,english]{revtex4}
\usepackage{amsmath,amssymb} % Enhanced mathematics
\usepackage[dvips]{graphicx} % Include figure files
\usepackage{graphics}
\usepackage{graphicx}
\usepackage{epsfig}
\usepackage{slashed}
\usepackage[english]{babel}
\usepackage{amsmath}
\usepackage{amssymb}
\usepackage{graphics}
\usepackage{graphicx}
\usepackage{epsfig}
\usepackage{slashed}
\usepackage{subfigure}	% Multiple figures on one plot
\usepackage[export]{adjustbox}
\newcommand\nn{\nonumber}
\newcommand\ba{\begin{eqnarray}}
\newcommand\ea{\end{eqnarray}}
\newcommand{\br}[1]{\left( #1 \right)}

\newcommand{\GeV}{~\mbox{GeV}}
\newcommand{\MeV}{~\mbox{MeV}}

\newcommand{\eV} {~\mbox{eV}}

\newcommand{\Tr}{\mbox{Tr}}

\begin{document}
\title{Total cross section of the process $e^+ + e^- \to \Sigma^+ + \bar{\Sigma}^-$
including the $D$-meson loop and three-gluon contributions}
\author{Azad I.~Ahmadov$^{a,b}$ \footnote{E-mail: ahmadov@theor.jinr.ru}}
\affiliation{$^{a}$ Institute of Physics,
Ministry of Science and Education, H.Javid avenue, 131, AZ1143 Baku, Azerbaijan}
\affiliation{$^{b}$ Bogoliubov Laboratory of Theoretical Physics,
JINR, Dubna, 141980 Russia}

\date{\today}

\begin{abstract}
In the present paper, we investigate the production of a baryon pair in $e^+e^-$ annihilation to study the structure of baryons.
To study the basic structure of the Standard Model, it is also necessary to consider the baryon-antibaryon pair production at
the electron-positron linear collider.
The production of a baryon pair in electron-positron annihilation provides a powerful tool at higher center-of-mass energies.
In this work, we present phenomenological results for $\Sigma^+ \bar {\Sigma}^-$ production in $e^+e^-$ collision at the BESIII Collider.
We investigate a hyperon pair produced in the reaction $e^+e^- \to \Sigma^+ \bar{\Sigma}^-$.
Two different contributions are considered: the $D$ meson loop and the three-gluon charmonium annihilation loop.
Taking into account the contributions of the $D$-meson loop and three-gluon loops and the interference of all diagrams,
we calculate the total cross section of the process $e^+e^- \to \Sigma^+ \bar{\Sigma}^-$.
With respect to a purely electromagnetic mechanism, large relative phases are generated for these contributions.
We obtain as a byproduct a fit of the electromagnetic form factor of the $\Sigma$ hyperon for a large momentum transferred region.
In this process, in addition to $\psi(3770)$ charmonium, we also take into account the contributions of other charmonium (like) states,
$\psi(4040)$, \,\,$\psi(4160)$, \,\,$Y(4230)$,\,\,$Y(4360)$,\,\,$\psi(4415)$, and $Y(4660)$.
The obtained results are in good agreement with experimental data.

\vspace*{0.5cm}

\noindent
\end{abstract}

\maketitle
% ======================================================================
\section{Introduction}
\label{Introduction}
% ======================================================================
It is understandable that the nucleons, as the lightest baryons, are the largest component of the observable matter in the Universe
and are shown to be nonpointlike particles \cite{A1,A2}.
In fact, the hyperons are the SU(3)-flavor-octet partners of the nucleons that contain one or more strange quarks and necessary
crucial additional dimensions to study nucleon structures \cite{A3,A4,A41,A42}.
It should be noted that to understand the strong interaction where quarks form hadrons is one of the most challenging questions
in contemporary physics.
It is known that electromagnetic form factors (EMFFs) are fundamental observables of baryons that are closely related to their
internal structure and dynamics.
That is, EMFFs, which parameterize the internal structure and dynamics of hadrons, are fundamental observables of hyperons and nucleons
for understanding the effects of Quantum Chromodynamics (QCD) in hadron resonances \cite{A5,A6,A7,A8}.

It must be noted that electromagnetic form factors describe well the changes at the point photon-hadron vertex due to the
internal structure and dynamics of hadrons.
Therefore, our knowledge of the electromagnetic form factors of the nucleon has significantly improved over the last years.
We can say that the electromagnetic form factors, can be investigated using the process of electron-positron interaction.
Really, it is possible to obtain valuable information about their structure from the electromagnetic form factors of
nucleons with high momentum transfer.

The pioneering work of Hofstaedter \cite{I1,I2}, who studied the electromagnetic structure of hadrons, still remains an
open and interesting area of research in high energy physics.
One of the main tools for studying the structure of the nucleon is the production of hadrons in electron-positron
interactions at high energies.

It should be noted that the first state of $Y$ named $Y(4260)$ was discovered by $BABAR$ in the process
$e^+e^- \to \pi^+ \pi^-J/{\psi}$ using the initial state radiation (ISR) method \cite{A9} and subsequently confirmed
by the collaborations of CLEO \cite{A10} and Belle \cite{A11} in the same process.

Like others, the vector ($J^{PC}=1^{--}$) charmoniumlike $Y$-family states, e.g., $Y(4360)$, were firstly
observed and confirmed in the process $e^+e^- \to (\gamma_{ISR})\pi^+\pi^- \psi(3686)$  by $BABAR$ and Belle experiments
\cite{A12,A13}. However, the nature of $Y(4360)$ has remained mysterious.

The $Y(4660)$ has been reported in ISR production in the process $e^+e^- \to (\gamma_{ISR}) \psi(2S)\pi^+\pi^-$
\cite{A121,A131}.
It was known that $Y(4660)$ was observed only in the Belle experiment \cite{A131} and, therefore, it is important
to confirm the existence of this state.

Also in the ISR study of the $e^+e^- \to \eta J/{\psi}$ process by the Belle experiment only the well-known
charmonium states $\psi(4040)$ and $\psi(4160)$ were discovered \cite{A16}.

It is known that the $\psi(3770)$ resonance is the lowest mass charmonium resonance above the open charm pair $D\bar{D}$
production threshold and that the $\psi(3770)$ resonance decays almost completely into pure $D\bar{D}$ \cite{A17}.
Namely, due to the discovery of many new states above the threshold of open charm, charmonium spectroscopy was revived.

It is known that the study of the physics of charm plays an important role in understanding the standard model.
It should be noted that experiments with electron-positron colliding beams have provided a number of interesting data on the
annihilation cross section in the vector meson region, as well as some preliminary data in the region of several GeV.
It is necessary to note that for studying the strong interaction in the transition region between the nonperturbative
and perturbative regimes of QCD hyperons are ideal probes.

The region in the center of mass with energies above the threshold for the production of the charmonium is of great interest
for the theory due to its richness of $c\bar {c}$ states, the properties of which have not been sufficiently studied.
We know that a lot of charmonia have been observed since 1974 \cite{A18,A19}, but this is not the end of the whole history of the
charmonium family, especially of observations of charmoniumlike XYZ states \cite{A20,A21}.

It is clear that the charmonium family group, by adding more states, may provide important information for understanding the
nonperturbative behavior of the strong interaction, which is intensively associated with color restriction in quantum chromodynamics (QCD).
The vector charmoniumlike $Y$ state plays a special role in the construction of the charmonium family.

To understand the strong interactions and the internal structure of hadrons, it is necessary to use timelike electromagnetic
form factors (EMFFs).
These form factors were measured using the process $e^+e^- \to B \bar {B}$ \cite{A22,A23,A24,A25,A26} (where $B$ denotes
the ground baryon state with spin 1/2).

We also would like to note that in experimental measurements, all charmonium states were observed below the threshold of a charmed
pair of quarks in mass ($2m_D$), and it was shown that their observed spectrum is consistent with the prediction of the model of the
potential of charm and anticharm \cite{A27}.

Two-particle baryon decays of vector ($J^{PC}=1^{--}$) charmonium (like) resonances provide a rich laboratory for
predictions of quantum chromodynamics \cite{A28,A29}.

In the near-threshold region, the cross sections of the processes $\bar {p} p\to e^+e^-$ and $e^+e^-\to\bar{p}p$ \cite{Haidenbauer} were analyzed
and the effective form factors (form factors $G_E$ and $G_M$) for the proton were used at energies close to the threshold of $\bar {p} p$.

For the process $e^+e^- \to \bar {p}p$, the authors of the work \cite{Bianconi} noted that there are deviations in the existence in the $BABAR$
data in timelike proton form factors from the pointlike behavior of proton-antiproton electromagnetic current.
Namely, in this paper $F_0$ + $F_{osc}$ was used in terms of the form factor where $F_0$ is expressed as the long-term trend of the form factor,
and $F_{osc}$ is a function that has the form exp(-Bp)cos(Cp) where $p$ is the relative momentum of the $\bar {p} p$ pair in the final state.

In the case of the Born approximation with a distorted wave, in the paper \cite{Qin} the effective EMFFs of a proton and a neutron in the
timelike region at the electron-positron annihilation into nucleon-antinucleon ($\bar {N} N$) pairs were investigated.

In the work \cite{Milstein}, the electromagnetic form factors $G_E$ and $G_M$ for protons and neutrons in the production
of $p\bar{p}$ and $n\bar{n}$ at the $e^+e^-$ annihilation near the threshold were used within the method of the effective optical potential,
which describes well the phases of scattering of $N\bar{N}$ and a sharp dependence of the cross sections of the formation of $N\bar{N}$.

In \cite{I3,I4}, it was suggested that in the process of electron-positron collisions for timelike regions, $q^2>0$
the electromagnetic form factors of hadrons can be studied by measuring the cross sections of hadron pair production.

In these experimental works \cite{A8,I5,I6,I8,I9,I10,I11,I13}, from the point of view of available data,
there is an electromagnetic form factor in a timelike region, which is consistent with naive rules for counting quarks and
the prediction of perturbative QCD (pQCD) at large $q^2$ \cite{I14,I15}.
Therefore, in these cases, to provide a unique opportunity to study the internal structure of hadrons and the electromagnetic
properties of hyperons, timelike form factors are used.

It should be noted that the data listed in \cite{I16}, which are obtained mainly from experiments on electron-positron and
proton-proton (proton-antiproton) interactions, determine today most of our knowledge about experimentally established nucleon resonances.
The charmonium resonance mass is present in the transition region between the perturbative and nonperturbative regimes of the $\psi(3770)$.
To gain knowledge about the structure in perturbative and nonperturbative strong interactions, it is necessary to study the strong and
hadronic decays of $\psi(3770)$ in this energy region \cite{I17}.

After analyzing these samples together with the data, it became possible to conduct a theoretical study taking into account the
interference of resonant and nonresonant amplitudes for the exclusive decay of $\psi(3770)$.
It is known that for several years BESIII has collected the largest sample of data on $e^+e^-$ collisions at an energy of 3.773 GeV.

It must be noted that charmonium states with $J^{PC}=1^{--}$, such as $J/\psi$, $\psi(3770)$,\,\, $\psi(4040)$, \,\,$\psi(4160)$,
\,\,$Y(4230)$,\,\,$Y(4360)$,\,\,$\psi(4415)$, $Y(4660)$ and others, are produced as
a result of electron-positron annihilation into a virtual photon at electron-positron colliders.

What follows is the decay process in these charmonium states, i.e., the decay into light hadrons via either a three-gluon process
($e^+e^-\to\psi \to ggg \to hadrons$) or a one-photon process ($e^+e^-\to \psi \to \gamma^* \to hadrons$).

Really, according to the Okubo-Zweig-Iizuki rule (OZI), it is expected that $\psi (3770)$, the lowest charmonium state
lying at $1^{--}$ above the threshold of $D\bar{D}$, will predominantly decay into the $D\bar{D}$ final states \cite{A17,I18,I20}.
To test a QCD prediction that can be understood in terms of the amplitudes of the quark distribution in hadron-hadron pairs and
the conservation of the total hadron helicity, it is necessary to study the process of $\psi(3770)$ production in $e^+e^-$ annihilation
and its subsequent decay into two hadrons.

In the theory of hadron physics, due to their wealth of $c\bar{c}$ states the $\psi(3770)$ \cite{I16} are of great interest,
which is one of those prominent structures in the hadron cross section.

It is worth noting that at the energy of 3.770 GeV the $\psi(3770)$ resonance have proven to be the only observed structure,
i.e. is the charmonium resonance with the lowest mass, which exceeds the open charm pair $D\bar{D}$ production threshold.

It can be expected that the resonance of $\psi(3770)$ will almost completely decay into pure $D\bar{D}$ \cite{A17}, and in the baryon sector
at the electron-positron collider, the production of a baryon-antibaryon pair can be tested by fundamental symmetries, in particular,
when the probability of the process increases due to the resonance such as the $J/\psi$ (or $\psi$)\cite{I22}.

It should be noted that the BESIII collaboration has until now conducted highly accurate studies of the possible threshold increase in the
$e^+e^- \to \Sigma^{\pm} \bar{\Sigma}^{\mp}$ \cite{I23},
$\Xi^-\bar{\Xi}^+$ \cite{I24,I24A} and $\Sigma^0 \bar{\Sigma}^0$ \cite{I25} processes and has also shown that the cross section does not
vanish near the threshold.

It can be understood that the threshold effect thus obtained will be useful for measuring the near-threshold pair generation of hyperons
in the production of $\Sigma^+\bar{\Sigma}^-$ in the $e^+e^-$ annihilation process, which was observed in the BESIII experiment \cite{I26}.

It is necessary to note that the process $e^+e^- \to \Sigma^+ \bar {\Sigma}^-$ was studied in detail by many authors
\cite{I26, I27,I28,I29,I30,I31,I32,I33}, and indeed, for the development of modern high-energy physics, a few experiments are very important.

Our aim in this paper is to study the characteristics of the $\Sigma^+ \bar {\Sigma}^-$ production in the $e^+e^-$ annihilation process,
including beyond the one-photon exchange also the exchange of several charmonium(like) states,
$\psi(3770),\, \psi(4040),\, \psi(4160),\,Y(4230),\, Y(4360),\, \psi(4415),\, \psi(4660)$.

\section{The process $e^+e^- \to \Sigma^+ \bar{\Sigma}^-$ in Born approximation \label{Born}}

In this section, we consider in the leading order the $\Sigma^+ \bar{\Sigma}^-$ production in the electron-positron annihilation process.
This process can be written as follows:
\ba
e^+(p_+) + e^-(p_-) \to \Sigma^+(q_+) + \bar{\Sigma}^-(q_-),
\label{A1}
\ea
where the quantities $p_+$ and $p_-$ are four moments of the $e^+$ and $e^-$ of particles,
and the quantities $q_+$ and $q_-$ are four moments of the $\Sigma^+$ and $\bar{\Sigma}^-$ particles.

The Mandelstam invariants for this process are defined by
\ba
s &=&(p_+ + p_-)^2 = (q_+ + q_-)^2;  \notag \\
t &=&(p_+ - q_+)^2 = (p_- - q_-)^2;  \notag \\
u &=&(p_+ - q_-)^2 = (p_- - q_+)^2,
\label{Mand1}
\ea
and from momentum conservation these satisfy
\ba
s + t + u = 2m_{e}^{2} + 2 M_{\Sigma}^2,
\label{3}
\ea
where $m_{e}$ and $M_{\Sigma}$ denote the masses of the electron and $\Sigma^{+,-}$ hyperon, respectively.
In the present work, we neglect the mass of the electron $m_e$.

The "master formula" of the differential cross section for the process \eqref{A1} in the Born approximation has the form:
\ba
d\sigma = \frac{1}{8 s} \sum_{\text{spins}} |{M}|^2 \, d\Phi_2,
\label{CrossSectionGeneralForm}
\ea
the summation is performed in the square of the matrix element over all possible spin states of the initial and final particles.

\begin{figure}
    \centering
    \includegraphics[width=0.40\textwidth]{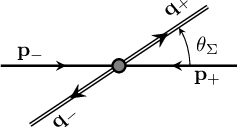}
    \caption{The definition of the scattering angle $\theta_\Sigma$ from (\ref{Phi2}) in
    the center-of-mass reference frame.}
    \label{fig.MomentaPosition}
\end{figure}

The phase volume element of the final particles $d\Phi_2$ can be written as follows:
\ba
&& d\Phi_2 = (2\pi)^4 \delta (p_+ + p_- - q_+ - q_-) \frac{d\bf{q_+}}{(2\pi)^3 2E_+}\frac{d\bf{q_-}}{(2\pi)^3 2E_-} = \nn\\
&&= \frac{1}{(2\pi)^2} \delta (p_+ + p_- - q_+ - q_-) \frac{d\bf{q_+}}{2E_+}\frac{d\bf{q_-}}{2E_-} =
    \frac{|\bf{p}|}{16 \pi^2 \sqrt{s}} \, d\Omega_\Sigma =\frac{\beta}{16 \pi} \, d\cos\theta_\Sigma,
    \label{Phi2}
\ea
where $d\Omega_\Sigma=d\phi_\Sigma \, d\cos\theta_\Sigma = 2\pi\, d\cos\theta_\Sigma$,
and $\phi_\Sigma$ and $\theta_\Sigma$ are the azimuthal and polar angles of the $\Sigma$-hyperon momentum
in the $e^+e^-$ rest frame,  that is $\theta_\Sigma$ is the angle between the directions of the momenta of the electron
$\bf{p_-}$ and $\Sigma$-hyperon $\bf {q}_+$ (Fig.~\ref{fig.MomentaPosition}).
We note that the modulus of three momenta of the final $\Sigma^+$-hyperon (or $\bar {\Sigma}^-$-hyperon) is defined in this case by the
$\delta$-function in the phase volume, i.e.,
\ba
&& |{\bf{p}}| \equiv |{\bf{q_+}}| = |{\bf{q_-}}| = \frac{\sqrt{s}}{2}\sqrt {1 - \frac{4M^2_{\Sigma}}{s}} = \frac{\sqrt{s}}{2} \beta, \\ \nn
&& \beta = \sqrt {1 - \frac{4M^2_{\Sigma}}{s}},
\ea
where $\beta$ is the $\Sigma^{+,-}$-hyperon velocity in the $e^+e^-$ center-of-mass frame,
$s$ is the square of the center-of-mass energy, and $M_{\Sigma}$ is the $\Sigma^{+,-}$-hyperon mass.

\begin{figure}
	\centering
    \subfigure[]{\includegraphics[width=0.40\textwidth]{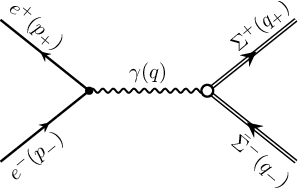}\label{BornDiagram}}
	\hspace{0.05\textwidth}
    \subfigure[]{\includegraphics[width=0.40\textwidth]{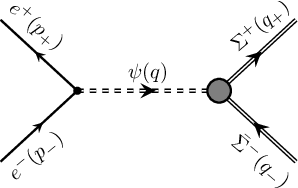}\label{PsiDiagram}}
    \caption{The Feynman diagrams for the hyperon pair production in the $e^+e^- \to \Sigma^+ \bar{\Sigma}^-$
    process corresponding via one-photon exchange (a) and the intermediate state $\psi(3770)$ charmonium (b).}
    \label{fig.TwoMechanisms}
\end{figure}

The four- moments of the initial-state leptons and the final-state $\Sigma^{+,-}$-hyperons can be parameterized as
\ba
p_- = \frac{\sqrt{s}}{2}(1,0,0,1),\,\,\,\,p_+ = \frac{\sqrt{s}}{2}(1,0,0,-1),  \nn \\
q_+ = \frac{\sqrt{s}}{2}(1,\beta \sin\theta_\Sigma,0, \beta \cos\theta_\Sigma),\,\,\,\,\, \nn
q_- = \frac{\sqrt{s}}{2}(1,-\beta \sin\theta_\Sigma,0,-\beta \cos\theta_\Sigma). \nn
\ea
The Feynman diagram for the $\Sigma^+ \bar{\Sigma}^-$ pair production in the electron-positron annihilation process
\eqref{A1} into a virtual photon in the Born approximation is presented in Fig.~\ref{BornDiagram}.

The process \eqref{A1} in the Born approximation, as shown in Fig.~\ref{BornDiagram}, is described by a cleanly electrodynamic diagram.
The amplitude for the process $e^+(p_+) + e^-(p_-) \to \Sigma^+(q_+) + \bar{\Sigma}^-(q_-)$, which corresponds to the Feynman diagrams,
in the Born approximation can be written in the form:
\ba
   {\mathcal M}_B = -\frac{e^2}{s} [\bar{v}(p_+) \gamma_\mu u(p_-)] \, [\bar{u}(q_+) \Gamma_\mu(q) v(q_-)],
    \label{BornAmplitude}
\ea
where the $e$ is the elementary electric charge, i.e., $e=\sqrt {4\pi\alpha}$, $\alpha \approx 1/137$ is the quantum electrodynamics (QED) fine structure constant
\cite{I16}, and $s$ is the squared total invariant mass of the electron-positron pair.

In this process, we have introduced form factors in order to describe the composite nature of hyperons (baryons).
In Fig.~\ref{BornDiagram} it is shown that for the annihilation process the current is $\gamma\Sigma^+\bar{\Sigma}^-$ $(\gamma B \bar{B})$
can be written using the Pauli form factors $F_1$ and $F_2$.
The vertex function $\Gamma_\mu(q)$ that describes the photon vertex with hyperons [Fig.~\ref{BornDiagram}] can be written in the following form:
\ba
	\Gamma_\mu(q) &= F_1(q^2) \gamma_\mu - \frac{F_2(q^2)}{4M_\Sigma} (\gamma_\mu \hat{q} - \hat{q} \gamma_\mu),
\label{FF}
\ea
where $M_{\Sigma}$ is the mass of the hyperon and $q$ is the momentum transferred.

The notation $\hat{q} = q_\nu\gamma^\nu$  will be used for $\hat{q}$.
It is necessary to note that in \eqref{FF} the functions $F_1(q^2)$ and $F_2(q^2)$ are the form factors of $\Sigma$-hyperons and
they are usually normalized in the following form:
$F_1(0)=0$ and $F_2(0)=\mu_{\Sigma}$, where $\mu_\Sigma$ is the $\Sigma$-hyperon anomalous magnetic moment.

However, the authors in \cite{I38} showed that this point behavior of the proton near the threshold is not so unambiguous, and therefore,
it is necessary to take into account the structure of these effects because the nontrivial structure of the baryon starts to manifest
itself already at a relatively small $q^2$.
Moreover, below will show that a comparison of the Born cross section with the experimental data will not be possible without the form factor.
It is worth noting that, in the timelike region there are few statistics and, therefore, large errors
in the experimental data, and at that time it is not possible to distinguish between the electric $G_E$ and magnetic $G_M$ form factors
in the experiment.

Therefore, considering this, we also, like many authors, use the approximation $|G_E| = |G_M|$, from which it follows that $F_2(q^2)=0$.
Therefore, here we use the effective form factor $F_1(q^2) = G(q^2)$, which was applied on the basis of QCD in \cite{I14,I40}.

For the square of the matrix element of \eqref{BornAmplitude} calculating the trace techniques and taking into account the form factor
$F_1(q^2) = G(q^2)$, we get the following expression:
\ba
    \sum_{spins} |{\mathcal M}_B|^2
    &=
    64 \pi^2 \alpha^2 |{G(s)}|^2 ( 2 - \beta^2 \sin^2\theta_\Sigma).
    \label{AmplitudeSquare}
\ea
Using formula \eqref{AmplitudeSquare} and taking into account the expressions for the phase volume in \eqref{Phi2}, we get the following
expression for differential cross section \eqref{CrossSectionGeneralForm}:
\ba
    d\sigma_B(s)=\frac{\pi\alpha^2 \beta}{2s}|{G(s)}|^2 (2-\beta^2 \sin^2\theta_\Sigma)\,\,d\cos\theta_\Sigma.
    \label{DTotalCrossSectionBorn}
\ea
After performing the integration of this expression \eqref{DTotalCrossSectionBorn} over all possible scattering angles
$d\cos\theta_\Sigma = \sin\theta_\Sigma\,d\theta_\Sigma$, we get the total cross section.
The integration limits for the angles $\theta_\Sigma$ are defined in the following form:
\ba
0 \leq \theta_\Sigma \leq \pi.  \nn
\label{theta}
\ea
After performing the integration over the angle $\theta_\Sigma$ for the total cross section in the Born approximation,
we obtain the following expression:
\ba
    \sigma_B(s)=\frac{2\pi\alpha^2}{3s} \beta (3-\beta^2) |G(s)|^2.
    \label{TotalCrossSectionBorn}
\ea
Here we use the form factor $G(s)$ for which the pQCD form from \cite{I14,I40} is applied, taking into account
the QCD running coupling constant $\alpha_s$ in the following form:
\ba
	G(s) = \frac{C}{s^2 \log^2\br{s/\Lambda_{QCD}^2}},
	\label{Formfactor}
\ea
where $C$ is a free parameter that fits the experimental data and $\Lambda_{QCD}$ is the QCD scale parameter.
We want to note that the constant $C$ should be fitted for the hyperon-antihyperon production according to specific
experimental data in the energy range of the corresponding experiment.

\begin{figure}
	\centering
	\includegraphics[width=0.55\textwidth]{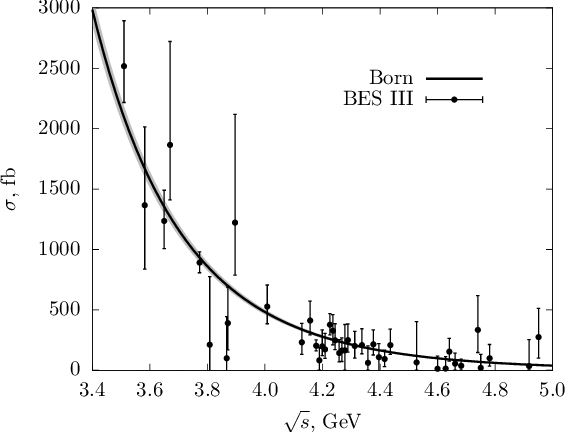}
    \caption{The total cross section of the process $e^+ e^-\to \Sigma^+ \bar{\Sigma}^-$ as a function of the center-of-mass energy $\sqrt {s}$.
The theoretical curve for the total cross section in the Born approximation (\ref{TotalCrossSectionBorn}) is denoted by black lines.
Due to errors in fitting the form factor constant (\ref{eq.C}), curve errors occur.}
    \label{WideRange}
\end{figure}

In the present work for the $\Sigma^+\bar{\Sigma}^-$ pair production in the $e^+e^-$ annihilation process \eqref{A1},
we fix this constant $C$ using the BESIII measurement \cite{I26}, and in Fig.~\ref{WideRange} we show the dependence of the total cross section as
a function of the center-of-mass energy $\sqrt {s}$.
When the value is $\Lambda_{QCD}$ = 300\,MeV in the Born cross section from (\ref{TotalCrossSectionBorn}) for the parameter $C$,
after fitting with respect to these data we obtain the following value:
\ba
	C = (66.59 \pm 1.4)~\GeV^4,
    \label{eq.C}
\ea
we use this value in further calculations for the electromagnetic form factor of the $\Sigma$-hyperon (\ref{Formfactor}).
Of course, it should be emphasized that the equation for the form factor $G(s)$ (\ref{Formfactor}), which has a constant $C$
from (\ref{eq.C}), applies to a relatively large transferred momentum $q^2$.
In this case, the Coulomb-like enhancement factor with many subtle features plays an important role \cite{Haidenbauer,Amoroso},
or manifests the wavelike nature of the hyperon of the stabilization after its emerging from the vacuum \cite{Egle};
therefore, it does not pretend to work near the threshold.

%

% =======================================================================
\section{The quarkonium $\psi(3770)$ intermediate state}
\label{sec.PsiIntermediateState}
% =======================================================================

We should like to note that the main task  of this work is to study the excitation effect of
the $\psi(3770)$ charmonium resonance in the process \eqref{A1}.
The total cross section of the dependence of the center-of-mass energy in the Born approximation \eqref{A1}, which is shown
in Fig.~\ref{WideRange}, including only the electromagnetic mechanism, cannot describe this delicate behavior near
the charmonium resonance $\psi(3770)$.
Therefore, we need to take into account in this region an additional contribution to the amplitude that appears from the
diagram with $\psi(3770)$ in the intermediate state (Fig.~\ref{PsiDiagram}) and which is enhanced by the Breit-Wigner propagator.
In this case, when exciting the charmonium resonance $\psi(3770)$ [$I^G (J^{PC})=0^-(1^{--})$] in the intermediate state,
we can calculate the contribution of the additional mechanism.
Thus, the total amplitude of the process \eqref{A1} is the sum of two matrix elements,
\ba
\mathcal {M} = \mathcal {M}_B + \mathcal {M}_{\psi},
\label{M}
\ea
where $\mathcal {M}_B$ denotes the amplitude, which is written in \eqref{BornAmplitude} in the Born approximation (Fig.~\ref{BornDiagram})
for the process \eqref{A1},
and $\mathcal {M}_{\psi}$ is the amplitude that takes into account the contribution enhanced by the Breit-Wigner factor [Fig.~\ref{PsiDiagram}]
for the intermediate state of the $\psi(3770)$ charmonium,
\ba
    M_\psi = \frac{1}{s-M_\psi^2+i M_\psi\,\Gamma_\psi} J^{e\bar{e}\to\psi}_\mu(q) \biggl(g^{\mu\nu} - \frac{q^\mu q^\nu}{M_\psi^2}\biggr)
    J^{\psi\to \Sigma^+\bar{\Sigma}^-}_\nu (q),
    \label{Mpsi}
\ea
where the quantity $M_{\psi}$ = 3773.7 \,MeV is the $\psi (3770)$ mass and the quantity $\Gamma_{\psi}$ = 27.2 \,MeV \cite{I16}
is the total decay width of the $\psi (3770)$ resonance.
The current $J^{e\bar{e}\to\psi}_\mu(q)$ determines the transition of an electron-positron pair into the resonance of the $\psi(3770)$ and
the current $J^{\psi\to \Sigma^+\bar{\Sigma}^-}_\nu (q)$ determines the transition of the resonance $\psi(3770)$ into the pair $\Sigma^+\bar{\Sigma}^-$, respectively.
Now, we take into account that the currents $J^{e\bar{e}\to\psi}_\mu(q)$ and $J^{\psi\to \Sigma^+\bar{\Sigma}^-}_\nu (q)$ in
\eqref{Mpsi} are conserved, that is, $q^{\mu}J^{e\bar{e}\to\psi}_\mu(q) = q^{\mu}J^{\psi\to \Sigma^+\bar{\Sigma}^-}_\mu (q) = 0.$
Using the work \cite{Ahmadov}, we accept that the vector current $J^{e\bar{e}\to\psi}_\mu(q)$ also has the same structure
as in the case of a photon according to \eqref{BornAmplitude} and can be written in the following form:
\ba
    J^{e\bar{e}\to\psi}_\mu(q) = g_e \, [\bar{v}(p_+) \gamma_\mu u(p_-)],
    \label{Jeepsi}
\ea
but only with a different constant $g_e = F_1^{\psi\to \Sigma^+ \bar{\Sigma}^-}(M_\psi^2)$, that depends on the charmonium mass
and is equal to the value of the form factor
$F_1^{\psi\to \Sigma^+ \bar{\Sigma}^-}(M_\psi^2)$ which is the value of the form factor for the vertex $\psi \to \Sigma^+\bar{\Sigma}^-$ on the $\psi(3770)$
mass shell [we note that here we accept the same approximation as in the Born case and assume
that $F_2^{\psi\to \Sigma^+ \bar{\Sigma}^-}(M_\psi^2) = 0]$.
We know from \cite{I16} the total decay width of $\psi\to e^+ e^-$, which is equal to $\Gamma_{\psi\to e^+e^-} = 261~\eV$,
and we can calculate this constant $g_e$,
\ba
    g_e= \sqrt{\frac{12\pi\Gamma_{\psi\to e^+e^-}}{M_\psi}} = 1.6 \cdot 10^{-3}.
\ea
It is shown \cite{Kuraev} that the imaginary part of the vertex $e\bar{e} \to \psi$ is small and
it is less than 10~\% of the real part. In this case, we neglect such a possible imaginary part and
we do not include this source of error when computing the statistical uncertainty of our final results.

After this, we can calculate the contribution of the charmonium in the intermediate state to the cross section.
Then, we can substitute the total matrix element from \eqref{M} into the general formula for the cross section
\eqref{CrossSectionGeneralForm}, and after that, we get the following expression for the total cross section:
\ba
	&\sigma \sim |\mathcal {M}|^2 = ||\mathcal {M}_B| + e^{i\phi} |\mathcal {M}_{\psi}||^2
	=\nn\\
	&\qquad = |\mathcal {M}_B|^2 + 2\cos\phi |\mathcal {M}_B| \cdot |\mathcal {M}_{\psi}| + |\mathcal {M}_{\psi}|^2
	\sim \sigma_B + \sigma_{int} + \sigma_\psi,
	\label{TotalCrossSection}
\ea
where $\phi$ is the relative phase between the Born amplitude $\mathcal {M_B}$ and the contribution of the intermediate
state $\psi(3770)$ of the charmonium $\mathcal {M}_\psi$.  \\
Now we need to calculate only the contribution of the charmonium $\sigma_\psi$ to the cross section and the contribution of the
interference of the Born amplitude in $\sigma_{int}$ with the charmonium amplitude in $\sigma_\psi$.
Further, taking into account the Born cross section $\sigma_B$ from \eqref{TotalCrossSectionBorn} and the interference
contribution $\sigma_{int}$ with the phase $\phi$, using \eqref{TotalCrossSection} we can calculate the total cross section
including both contributions, and after this we can evaluate $\sigma_\psi$ in the following manner:
\ba
	\sigma_\psi = \biggl(\frac{\sigma_{int}}{2 \cos\phi \, \sqrt{\sigma_B}} \biggr)^2.
\label{sigmapsi}
\ea
Now we have to calculate the interference of the Born contribution $\mathcal{M_B}$ with the contribution of the intermediate
state of the $\psi(3770)$ charmonium $\mathcal {M}_\psi$.
Based on the general formula \eqref{CrossSectionGeneralForm}, the interference contribution to the cross section of this
process can be written in the following standard form:
\ba
    d\sigma_{int} = \frac{1}{8s} \sum_{\text{spins}} 2\,\mbox{Re}[\mathcal {M}_B^+ \mathcal {M}_\psi] \, d\Phi_2,
    \label{dsigmaint}
\ea
After integrating in \eqref{dsigmaint} over the phase space of the final particles, we obtain an expression for the contribution
to the total cross section in the following form:
\ba
    \sigma_{int}(s)
    &=
    \frac{1}{4 s^2} \mbox{Re}\biggr\{
    	\frac{\sum_{s}
    	(J^{e\bar{e}\to\gamma}_\mu)^* J^{e\bar{e}\to\psi}_\nu}
    	{s-M_\psi^2+i M_\psi\,\Gamma_\psi} \cdot
    	\sum_{s'}
    	\int d\Phi_2
    	(J_{\gamma\to \Sigma^+\bar{\Sigma}^-}^\mu)^* J_{\psi\to \Sigma^+\bar{\Sigma}^-}^\nu \biggr\},
    \label{sigmaint}
\ea
where $\sum_s$ is the summation over the initial particles spin states and $\sum_{s'}$ is the summation over the
spin states for the final particles.
Here we can use the method of invariant integration over the total volume of the final phase, and after this procedure
for the second term in \eqref{sigmaint} we obtain the following expression:
\ba
	&\sum_{s'} \int d\Phi_2	(J_{\gamma\to \Sigma^+ \bar{\Sigma}^-}^\mu)^* J_{\psi\to \Sigma^+ \bar{\Sigma}^-}^\nu
	= \frac{1}{3} \biggl( g^{\mu\nu} - \frac{q^\mu q^\nu}{q^2} \biggr)
	\sum_{s'} \int d\Phi_2 	\br{J_{\gamma\to \Sigma^+ \bar{\Sigma}^-}^\alpha}^* J^{\psi \to \Sigma^+ \bar{\Sigma}^-}_\alpha.
\label{JJ}
\ea
If we use the explicit expressions for the lepton currents $J^{e\bar{e} \to \gamma}_\mu$ from \eqref{BornAmplitude} and
$J^{e\bar{e}\to\psi}_\nu$ from \eqref{Jeepsi} and apply the law of conservation of currents, we can calculate
\ba
&&	\sum_{s}(J^{e\bar{e} \to \gamma}_\mu)^* J_{e\bar{e} \to \psi}^\mu
 = -e g_e \sum_{s} [\bar{u}(p_-) \gamma_\mu(q) v(p_+)] [\bar{v}(p_+) \gamma^{\mu} u(p_-)] \approx  \nn \\
&&  \approx
	-e g_e \, \Tr[\hat{p_-} \gamma_\mu \hat{p_+} \gamma^\mu] \approx 4 \, e \, g_e s.
\label{JJe}
\ea
If we use the explicit expression of the two-particle phase volume of the final particles from \eqref{Phi2} and the invariant integration
method from \eqref{JJ} and substitute only the obtained result \eqref{JJe} in \eqref{sigmaint} we obtain the expression for the interference
contribution to the total cross section in the following form:
\ba
    \sigma_{int}(s)
    = \frac{e g_e \beta}{48 \pi s}
       \mbox{Re}\biggl\{\frac{1}{s-M_\psi^2+i M_\psi\, \Gamma_\psi}
    	\int\limits_{-1}^1 d\cos\theta_{\Sigma^+} \sum_{s'}
    	(J_{\gamma\to \Sigma^+ \bar{\Sigma}^-}^\alpha)^* J^{\psi \to \Sigma^+ \bar{\Sigma}^-}_\alpha \biggr\}.
    \label{TotalCrossSectionInterference}
\ea
In this formula \eqref{TotalCrossSectionInterference}, the subintegral expression contains the dynamics of the transformation
of the charmonium into the $\Sigma^+\bar{\Sigma}^-$ pair, and therefore, it can be written in a separate form
\ba
	S_i(s)
    = \frac{e g_e \beta}{48 \pi s}
    	\int\limits_{-1}^1 d\cos\theta_{\Sigma^+} \sum_{s'}
    	(J_{\gamma\to \Sigma^+\bar{\Sigma}^-}^\alpha)^* J^{\psi\to \Sigma^+\bar{\Sigma}^-}_\alpha.
    \label{Si}
\ea

Thus, if we substitute formula \eqref{Si} into formula \eqref{TotalCrossSectionInterference},
then the expression for the interference contribution to the total cross section can be written in the following form:
\ba
	\sigma_{int}(s)	=
	\mbox{Re}\biggl(\frac{S_i(s)}{s-M_\psi^2+i M_\psi\, \Gamma_\psi}\biggr),
	\label{SigmaIntViaSi}
\ea
The subscript index $i$ in \eqref{Si} and \eqref{SigmaIntViaSi} denotes different mechanisms of this transformation.
For example, it is shown in Fig.~\ref{fig.DD} that the Feynman diagram describes quite possibly the usual OZI-permitted mechanism
through the $D$-meson loop.
It should be noted that the charmonium mass $\psi (3770)$ exceeds the threshold for the production of the $D\bar{D}$ pair,
so it is natural to expect that the $D$-meson loop will be the main mechanism in this process.
However, since the threshold excess is very small ($M_{\psi} -2 M_D \approx$ 39 MeV, i.e., relative to the characteristic energies
in the problem, this value is about 1\%), we expect this contribution to be small, and, therefore, it is also necessary to consider
other possible mechanisms.
We assume that a significant contribution will be made by the OZI-violating mechanism due to the three-gluon annihilation of the
charmonium to the $\Sigma^+\bar{\Sigma}^-$ pair, which is shown in Fig.~\ref{fig.3G}.

In order to reconstruct the total cross section, we use the procedure described in \cite{Bystritskiy} [Eqs.~(15) and (16)]
and with the help of the interference contribution (\ref{SigmaIntViaSi}) with the total relative phase between the Born amplitude
$\mathcal {M}_B$ and the amplitude with the charmonium contribution $\mathcal {M}_\psi$.

% =======================================================================
\section{The \textit{D}-meson loop mechanism}
\label{sec.DMesonLoopMechanism}
% =======================================================================

\begin{figure}
	\centering
    \includegraphics[width=0.50\textwidth]{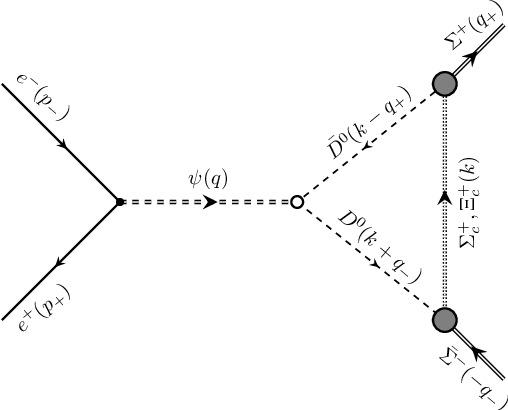}
    \caption{The Feynman diagram to the $\Sigma^+\bar{\Sigma}^-$ production in charmonium decays via $D$-meson
    loop mechanism in the process $e^+e^- \to \Sigma^+\bar{\Sigma}^-$ at the one-loop level.}
    \label{fig.DD}
\end{figure}

Figure~\ref{fig.DD} presents the Feynman diagram for the production of $\Sigma^+ \bar{\Sigma}^-$ via the $D$-meson loop,
which in this section, and according to this mechanism, we calculate the contribution of the intermediate charmonium with
the transition to the final state of the $\Sigma^+ \bar{\Sigma}^-$ hyperon pair through the $D$-meson loop.
For the calculation of the contribution of the $D$ meson loop to the cross section, we first need to calculate the value of
$S_D$ from \eqref{Si}, because it enters into \eqref{SigmaIntViaSi}.
To extract the quantity $S_D$ from $\mathcal {M}_D$, we need to construct the amplitude $\mathcal {M}_D$ corresponding to Fig.~\ref{fig.DD}.
Thus, according to Feynman's rules, the contribution of the $D$-meson loop to the amplitude corresponding to the diagram depicted in
Fig.~\ref{fig.DD}, we can write in the following form:
\ba
&&	\mathcal{M}_D =
    \frac{g_e}{16 \pi^2}
    \frac{[\bar{v}(p_+) \gamma_\mu u(p_-)]}{q^2-M_\psi^2+i M_\psi\,\Gamma_\psi}
    \cdot
    \int\frac{dk}{i \pi^2}
    \frac{[\bar{u}(q_+) \gamma_5 (\hat{k} + M_\Xi) \gamma_5 v(q_-)] (2 k + q_- - q_+)^\mu}{(k^2 - M_\Xi^2)((k-q_+)^2 - M_D^2) ((k+q_-)^2 - M_D^2)}
    \times\nn\\
&& \times G_{\psi D\bar{D}}(q^2,(k+q_-)^2,(k-q_+)^2) G_{\Xi D \Sigma}(k^2,(k-q_+)^2) G_{\Xi D \Sigma}(k^2,(k+q_-)^2),
    \label{AmplitudePsiDD}
\ea
where the quantities $M_\psi$, $M_D$, and $M_\Xi$ are the masses of the $\psi (3770)$-charmonium, $D$-meson and $\Xi$-hyperon, respectively.
In this case, we use the following dependence of the form factor on the  $\psi D\bar{D}$ vertex in this form:
\ba
	G_{\psi D\bar{D}} (s,M_D^2,M_D^2) =
	g_{\psi D\bar{D}} \, \frac{M_\psi^2}{s} \, \frac{\log (M_\psi^2/\Lambda_D^2)}{\log (s/\Lambda_D^2)},
	\label{PsiDDFormfactor}
\ea
where the scale $\Lambda_D$ is fixed on the characteristic value of the process $\Lambda_D = 2 M_D$.
After comparing this expression with the general form of the amplitude from \eqref{Mpsi}, we obtain for the current
$J^{\psi\to \Sigma^+\bar{\Sigma}^-}_\nu (q)$ the following expression:
\ba
&& J^{\psi\to \Sigma^+\bar{\Sigma}^-}_\mu (q) = \frac{1}{16 \pi^2}
\int\frac{dk}{i \pi^2}
    \frac{[\bar{u}(q_+) \gamma_5 (\hat{k} + M_\Xi) \gamma_5 v(q_-)] (2 k + q_- - q_+)^\mu}{(k^2 - M_\Xi^2)((k-q_+)^2 - M_D^2) ((k+q_-)^2 - M_D^2)}
   \times\nn\\
&& \times G_{\psi D\bar{D}}(q^2,(k+q_-)^2,(k-q_+)^2) G_{\Xi D \Sigma}(k^2,(k-q_+)^2) G_{\Xi D \Sigma}(k^2,(k+q_-)^2),
\label{JDD}
\ea
and it can be inserted into \eqref{Si}.
Thus, for $S_D$ \eqref{Si} we obtain the expression in the following form:
\ba
&&	S_D(s) = \frac{\alpha g_e \beta G(s)}{48 \pi^2 s} \int\frac{dk}{i \pi^2}
    \frac{Tr D(s, k^2)}{(k^2 - M_\Xi^2)((k-q_+)^2 - M_D^2) ((k+q_-)^2 - M_D^2)}
   \times\nn\\
&&  \times G_{\psi D\bar{D}}(q^2,(k+q_-)^2,(k-q_+)^2) G_{\Xi D \Sigma}(k^2,(k-q_+)^2) G_{\Xi D \Sigma}(k^2,(k+q_-)^2 = \nn \\
&&  \alpha_{D}(s) Z_D(s).
   \label{SDG}
\ea
Here the expressions for $\alpha_{D}(s)$ and $Z_D(s)$ have the following forms:
\ba
&& \alpha_{D}(s) = \frac{\alpha g_e \beta G(s)}{48 \pi^2}, \nn \\
&& Z_D(s) = \frac{1}{s} \int\frac{dk}{i \pi^2}
    \frac{Tr D(s, k^2)}{(k^2 - M_\Xi^2)((k-q_+)^2 - M_D^2) ((k+q_-)^2 - M_D^2)}
   \times\nn\\
&&  \times G_{\psi D\bar{D}}(q^2,(k+q_-)^2,(k-q_+)^2) G_{\Xi D \Sigma}(k^2,(k-q_+)^2) G_{\Xi D \Sigma}(k^2,(k+q_-)^2.
  \label{ZD}
\ea
where $Tr D(s, k^2)$ is the trace of the $\gamma$-matrices over the baryon line have the following form:
\vspace*{-0.5cm}
\ba
&&	Tr D(s, k^2) =
     Tr [(\hat {q_+}+M_\Sigma) \gamma_5 (\hat {k}+M_\Xi) \gamma_5 (\hat {q_-}-M_\Sigma) (\hat {k} - M_\Sigma)]
    =\nn\\
&&  = 2\left((k^2)^2 + k^2 (s - 2 (M_D^2 + M_\Sigma M_\Xi)) - s M_\Sigma M_\Xi + c_D  \right),
    \label{SpDExplicit}
\ea
where $c_D$ has the following expression:
\ba
&c_D
    =
    M_D^4 + 2 M_\Sigma M_\Xi M_D^2 + 2 M_\Xi M_\Sigma^3 - M_\Sigma^4.
    \label{cD}
\ea
For the vertices $\psi \to D\bar{D}$ and $D \to \Xi \Sigma$, the quantities $G_{\psi D\bar{D}}$ and $G_{\Xi D \Sigma}$
in formula \eqref{SDG} are the form factors \cite{I9,I25,I26,I30}. \\
Thus, using the Cutkosky cutting rules \cite{Cutkosky}, we can replace the $D$-meson propagators in \eqref{ZD} through
the $\delta$-functions, then we obtain the imaginary part of this quantity from $Z_D(s)$,
\ba
&&	2i \, \mbox{Im} \, Z_D(s)
	=\frac{(-2\pi i)^2}{s}
    \int\frac{dk}{i \pi^2}
    \frac{Tr D(s, k^2)}{k^2 - M_\Xi^2}
    G_{\psi D\bar{D}}(s,(k+q_-)^2,(k-q_+)^2)
    \times\nn\\
&&   \times
    G_{\Xi D \Sigma}(k^2,(k-q_+)^2)~G_{\Xi D \Sigma}(k^2,(k+q_-)^2)
    \delta ((k+q_-)^2 - M_D^2)~\delta ((k-q_+)^2 - M_D^2)
    \times\nn\\
&&    \times
    \theta ((k+q_-)_0)~\theta (-(k-q_+)_0).
    \label{ImZD0}
\ea
After performing the replacement of these two $\delta$-functions in \eqref{ImZD0} and implementing cyclic integrations,
we get for the imaginary part of this quantity the final expression as follows:
\ba
&&	\mbox{Im} \, Z_D\br{s}=
	-\frac{2\pi}{s^{3/2}}
	G_{\psi D\bar{D}}(s,M_D^2,M_D^2)
	\int\limits_{C_k^{(1)}}^1 \frac{dC_k}{\sqrt{D_1}}
    \sum_{i=1,2}
    \frac{k_{(i)}^2} {k_{(i)}^2 + M_\Xi^2}~
	\times\nn\\
&& \quad\times
    Tr D(s,-k_{(i)}^2)~G_{\Xi D \Sigma}^2(-k_{(i)}^2,M_D^2),
    \qquad
	s > 4M_D^2,
    \label{ImZD}
\ea
where $C_k = \cos\theta_k$ is the polar angle.
We can define the quantities $k_{(i)}$, $D_1$ and $C_k^{(1)}$ in this  form:
\ba
	k_{(1,2)} = \frac{1}{2} (\sqrt{s} \, \beta \, C_k \pm \sqrt{ D_1 }),
	\,\,\,
	D_1 = s \, \beta^2 \, C_k^2 - 4 (M_D^2 - M_{\Sigma}^2),
    \,\,\,
    C_k^{(1,2)} = \pm \frac{2}{\beta} \sqrt{\frac{M_D^2 - M_{\Sigma}^2}{s}}.
\ea
For the calculation of $\mbox{Im}\, Z_D$ from (\ref{ImZD}), an explicit expression of the form factor must be considered.
Note that in this case, for the $\psi\to D\bar{D}$ vertex, only the dependence on the virtuality of the charmonium $q^2=s$ is needed.
Thus, in this region we are interested in the dependence of the function $G_{\psi D\bar{D}}(s,M_D^2,M_D^2)$.
Considering that the legs of the $D$ meson are on the mass shell and then using expression (\ref{ImZD}),
we calculate the imaginary part of $Z_D$.

It is necessary to note that the details of calculation of the quantity $Z_D\br{s}$ can be found in \cite{Ahmadov,Bystritskiy}.
In the present work, we technically calculate the imaginary part of this value, and then using the dispersion relation with
one subtraction at $q^2=0$, we restore its real part.
It should be noted that the vertex $\psi \to \Sigma^+ \bar{\Sigma}^-$ at $q^2=0$ is equal to zero, because $\Sigma$ is a hyperon
(which is a $uds$ quark state) and has no open charm.

At the decay of $\psi(3770)$ into $D\bar{D}$ of this vertex function $G_{\psi D\bar{D}}(s,M_D^2,M_D^2)$ the normalization should be fixed.
In order to fix the functions, it is convenient to use the quantity of the decay width of $\psi \to D \bar{D}$,
\vspace*{-0.5cm}
\ba
g_{\psi D\bar{D}} \equiv G_{\psi D\bar{D}}(M_\psi^2, M_D^2, M_D^2).
\ea
In order to get the vertex $\psi \to D\bar{D}$ with the only dependence on the charmonium virtuality $q^2 = s$,
we can cut the diagram with the $D$-meson propagators, whose $D$-meson legs are on the mass shell.

After calculating the decay width of charmonium $\psi \to D \bar{D}$, we find the value of the constant $g_{\psi D\bar{D}}$.
Using the standard formula, we calculate the decay widths and get an expression for the total decay width as follows:
\ba
\Gamma_{\psi D\bar{D}} = \frac{g_{\psi D\bar{D}}^2 M_{\psi}\beta_{D}^3}{48 \pi}.
\ea
We know the experimental value for the width of the charmonium decay $\Gamma_{\psi D\bar{D}} = 25 \,MeV$ \cite{I16};
therefore, we can find the value of the constant $g_{\psi D\bar{D}}$ and obtain
\ba
g_{\psi D\bar{D}} = 4 \sqrt{\frac{3 \pi \Gamma_{\psi D\bar{D}}}{M_{\psi}\beta_{D}^3}} \approx 18.4,
\ea
where $\beta_D = \sqrt{1 - 4M_D^2/M_\psi^2}$ is the velocity of the $D$-meson in this decay.

Now we want to consider the function $G_{\Xi D \Sigma} (k^2,p^2)$ from (\ref{ZD}).
Since $k^2 < 0$, in the $t$-channel the only dependence in the imaginary part of $Z_D$ is the off-mass shell of the $\Xi$ baryon.

We would like to note that in \cite{Ahmadov,Bystritskiy,BA} we used a constant for the $\Lambda D P$-vertex based on the results of \cite{Reinders,Navarra}.
But, in this paper we use for $\Xi D \Sigma$-vertex the constant that corresponds to the results of \cite{Reinders,Choe}.

It should be noted that for the function $G_{\Xi D \Sigma}$ the $SU(4)$ symmetry leads us to the same result,
\ba
	G_{\Xi D \Sigma}(k^2, M_D^2) = \frac{f_D \, g_{\Xi D \Sigma}}{m_u + m_c},
	\qquad
	k^2 < 0,
\ea
where $f_D \approx 180~\MeV$ and
\vspace*{-0.5cm}
\ba
	g_{\Xi D \Sigma} \approx g_{K \Sigma \Xi} = -7.02.
	\label{gLambdaDXi}
\ea
For quark masses of $m_u$  and $m_c$, the following values are used: $m_u\approx 280~\MeV$ and $m_c = 1.27~\GeV$\cite{I16}.

% ========================================================================================================================
\section{The three-gluon mechanism}
\label{sec.ThreeGluonMechanism}
% ========================================================================================================================

\begin{figure}
	\centering
    \includegraphics[width=0.50\textwidth]{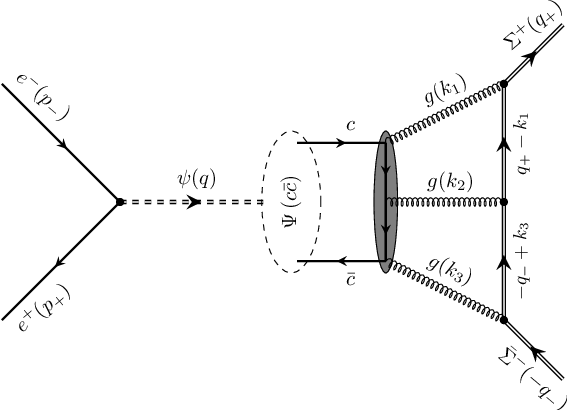}
    \caption{The Feynman diagram of the $\Sigma^+\bar{\Sigma}^-$ production in charmonium decays via three-gluon
    mechanism in the process $e^+e^- \to \Sigma^+\bar{\Sigma}^-$.}
    \label{fig.3G}
\end{figure}

In this section, we consider the contribution of the intermediate charmonium with the transition to the produced
$\Sigma^+ \bar{\Sigma}^-$-pair in the final state via the three-gluon annihilation mechanism.
To determine the contribution to the cross section, we must calculate the quantity of $S_{3g}$ from \eqref{Si},
which enters into \eqref{SigmaIntViaSi}.
Figure~\ref{fig.3G} shows the corresponding Feynman diagram for the $\Sigma^+ \bar{\Sigma}^-$ production through the three-gluon mechanism.
Note that the production of a $\Sigma^+ \bar{\Sigma}^-$ pair in the $e^+e^-$
annihilation process via the three-gluon mechanism was considered in \cite{Ahmadov,Bystritskiy}; it was revised and some typos and
minor errors were corrected.
In the present paper, we simply apply this mechanism to the production of the $\Sigma^+\bar{\Sigma}^-$ hyperon pair in the $e^+e^-$
annihilation process through a three-gluon exchange.

Thus, we can write the following contribution to the quantity of $S_{3g}$ from \eqref{SigmaIntViaSi} according to the corresponding
Feynman diagram (Fig.~\ref{fig.3G}), [coinciding with Eqs. (16) and (17) from \cite{Ahmadov}] in the interference of the charmonium
state with the Born amplitude [see (\ref{Si})]:
\ba
    S_{3g}(s) = \alpha_{3g} (s) \, Z_{3g}(s),
    \label{S3g}
\ea
where
\ba
&&    \alpha_{3g} (s) = \frac{\alpha\, \alpha_s^3}{2^3 \, 3} g_e \, g_{col} \, \phi \, \beta \, G (s) \, G_\psi(s),
    \label{alpha3g}
    \\
&&    Z_{3g} (s)= \frac{4}{\pi^5 s}\int \frac{dk_1}{k_1^2}\frac{dk_2}{k_2^2}\frac{dk_3}{k_3^2}
    \frac{Tr3g ~ \delta (q-k_1-k_2-k_3)}{((q_+ - k_1)^2 - M_\Sigma^2) ((q_- - k_3)^2 - M_\Sigma^2)},
    \label{Z3g}
\ea
where $g_{col}$ is the color factor, $g_{col} (1/4) = \left<\Sigma| d^{ijk} ~ T^i T^j T^k |\Sigma \right> = 15/2$.
The $Tr3g$ in \eqref{Z3g} is the product of traces over the lines of the $\Sigma$-hyperon and $c$-quark,
\ba
&&	Tr3g = \Tr [\hat{Q}_{\alpha\beta\gamma} (\hat {k}_c + m_c) \gamma^\mu (\hat {k}_{\bar{c}} - m_c)] \cdot
    \Tr\left[(\hat {q}_+ +M_\Sigma) \gamma^\alpha (\hat{q}_+ - \hat {k}_1 + M_\Sigma) \gamma^\beta \right.
	\times     \nn \\
&& \qquad\times \left. (-\hat {q}_- + \hat{k}_3 + M_\Sigma) \gamma^\gamma (\hat {q}_- -M_\Sigma) \gamma_\mu \right],
   	\nn
\ea
where
\ba
&&    \hat{Q}_{\alpha\beta\gamma}  =
    \frac{\gamma_\gamma (-\hat {k}_{\bar{c}} + \hat {k}_3 + m_c) \gamma_\beta (\hat {k}_c - \hat {k}_1 + m_c) \gamma_\alpha}
    {((k_{\bar{c}}-k_3)^2 - m_c^2) ((k_c-k_1)^2 - m_c^2)} +
     \frac{\gamma_\beta (-\hat {k}_{\bar{c}} + \hat {k}_2 + m_c) \gamma_\gamma (\hat {k}_c - \hat {k}_1 + m_c) \gamma_\alpha}
    {((k_{\bar{c}}-k_2)^2 - m_c^2) ((k_c-k_1)^2 - m_c^2)} +  \nn \\
&& + \frac{\gamma_\gamma (-\hat {k}_{\bar{c}} + \hat {k}_3 + m_c) \gamma_\alpha (\hat {k}_c - \hat {k}_2 + m_c) \gamma_\beta}
    {((k_{\bar{c}}-k_3)^2 - m_c^2) ((k_c-k_2)^2 - m_c^2)} +
    \frac{\gamma_\alpha (-\hat {k}_{\bar{c}} + \hat {k}_1 + m_c) \gamma_\gamma (\hat {k}_c - \hat {k}_2 + m_c) \gamma_\beta}
    {((k_{\bar{c}}-k_1)^2 - m_c^2) ((k_c-k_2)^2 - m_c^2)} +  \nn \\
&& + \frac{\gamma_\beta (-\hat {k}_{\bar{c}} + \hat {k}_2 + m_c) \gamma_\alpha (\hat {k}_c - \hat {k}_3 + m_c) \gamma_\gamma}
    {((k_{\bar{c}}-k_2)^2 - m_c^2) ((k_c-k_3)^2 - m_c^2)} +
    \frac{\gamma_\alpha (-\hat {k}_{\bar{c}} + \hat {k}_1 + m_c) \gamma_\beta (\hat {k}_c - \hat {k}_3 + m_c) \gamma_\gamma}
    {((k_{\bar{c}}-k_1)^2 - m_c^2) ((k_c-k_3)^2 - m_c^2)}. \,\,\,\,\,\,\,\,\,\,\,\,
    \label{DefinitionHatQ}
\ea
We will use equation (5) from (\cite{Chiang}) to properly normalize the color wave function.
The color wave function normalized to unity should be in the following form:
\ba
\frac{1}{\sqrt{3}}\bar{q}_i q_i = \frac{1}{\sqrt{3}} (\bar{q}_1 q_1 + \bar{q}_2 q_2 + \bar{q}_3 q_3). \nn
\ea
The factor $1/\sqrt{3}$ grants the correct normalization of this wave function.

It is necessary to note that the parameter $\phi$ in (\ref{alpha3g}) is related to the charmonium wave function and can be written in the following form:
\ba
	\phi = \frac{|\psi (\bf {r}=\bf{0})|}{M_\psi^{3/2}} = \frac{\alpha_s^{3/2}}{3\sqrt{3\pi}}.
	\label{eq.phi}
\ea
This quantity is derived from the $\psi \to 3 \,g$ decay rate on the mass shell.
It should be noted that according to Eqs.\eqref{alpha3g} and \eqref{eq.phi} this three-gluon mechanism is very sensitive to this
value of $\phi$ since at the charmonium scale (for $s \sim M_c^2$) it depends on its value to a rather high degree.
We want to note that in this paper, in our calculations, we use the value $\alpha_s(M_c) = 0.28$, which is expected by the QCD
evolution of $\alpha_s$ from the $b$-quark scale to the $c$-quark scale.
We note that this value differs from the value for the charmonium of $J/\psi$, where a much smaller value of the parameter
$\alpha_s(M_c) = 0.19$ was used \cite{Chiang}.

Here one of the most important corrections concerns the final state of $\Sigma^+ \bar {\Sigma}^-$.
During the process, after the decays of $\psi(3770)$, three gluons are obtained that produce three quark-antiquark pairs,
and further in the final state, $\Sigma^+ \bar {\Sigma}^-$ pairs are formed.
Now we need to reproduce the absolute value of the cross section, in order to implement this mechanism.
Thus, the transition of three gluons (that is, with the total angular momentum equal to $1$) into the final pair of
$\Sigma^+ \bar {\Sigma}^-$, is one of the purposes of this mechanism.

According to \cite{Bystritskiy,BA}, we assume that this mechanism has much in common with calculation in the
timelike region with the production of proton-antiproton and $\Lambda \bar {\Lambda}$ pairs from a photon.
The Formula (\ref{alpha3g}) includes the factor $G_\psi(s)$, which we accept as a form factor, since this factor describes
the mechanism of transition of three gluons into the final pair $\Sigma^+ \bar{\Sigma}^-$.
Hence, we can insert an additional form factor, analogical to (\ref{Formfactor}) into (\ref{alpha3g}), but with a different
value of the parameter $C_\psi$
\ba
	|G_\psi(s)| = \frac{C_\psi}{s^2 \log^2 (s/\Lambda^2)}.
	\label{FormfactorPsi}
\ea
Since gluons do not feel the flavor of quarks in the final baryons, in this paper, when calculating the constant $C_\psi$,
we use the same value as in the case of the production of a proton-antiproton pair and a pair $\Sigma^+ \bar{\Sigma}^-$ \cite{Bystritskiy,BA},
\ba
	C_\psi = (45 \pm 9)~\GeV^4.
	\label{eq.CPsi}
\ea
For restoring the real part of $Z_D$ from (\ref{ZD}) and $Z_{3g}$ from (\ref{Z3g}), we use the dispersion relation technique.
To do this, all the details of these calculations are described in \cite{Ahmadov}.
In this case, we get the following expression for the real part of $Z_D$ and $Z_{3g}$:
\ba
	\mbox{Re} \, Z_i(\beta)	=
	\frac{1}{\pi} \left\{
		\mbox{Im} \, Z_i(\beta) \log \left|\frac{1-\beta^2}{\beta_{\text{min}}^2 - \beta^2}\right|
		+ \int\limits_{\beta_{\text{min}}}^1 \frac{2 \beta_1 d\beta_1}{\beta_1^2 - \beta^2}
		[\mbox{Im} \, Z_i(\beta_1) - \mbox{Im} \, Z_i(\beta)]\right\}.
	\label{DispersionRelation}
\ea
We want to note that the imaginary part $\mbox{Im} \, Z_D(\beta)$ of (\ref{ImZD}) for the $D$-meson loop contribution is
non zero above the threshold ($s > 4M_D^2$); thus the integration over the lower limit in {(\ref{DispersionRelation})
is equal to $\beta_{\text{min}} = \sqrt{1 - M_{\Sigma}^2/M_D^2}$.
Note that for the three-gluon contribution the threshold of the imaginary part $\mbox{Im} \, Z_{3g}(\beta)$ with the
reaction threshold, i.e., $s_{\text{min}} = 4M_{\Sigma}^2$, and that is why we can assume that the lower limit of
integration will be equal to $\beta_{\text{min}} = 0$.

% =================================================================================================
\section{The Numerical results}
\label{sec.Numerical}
% =================================================================================================

In this section, we want to present the numerical results of the distribution of the total cross section from the total energy
at electron-positron collisions at BESIII energies \cite{I26} and, according to our calculation, the functions of $Z_D (s)$ and
$Z_{3g} (s)$ of the total energy $\sqrt {s}$ in the range starting from the reaction threshold $s=4M_{\Sigma}^2$ up to 4.951 GeV.
We compare the theoretical results we have obtained with the experimental data of BESIII.
In this paper, using formula \eqref{TotalCrossSectionBorn}, we calculate the total cross section for the processes of
$\Sigma^+ \bar{\Sigma}^-$ pair production as a function of the collider center-of-mass energy $\sqrt s$ in the range from 3.51 GeV to 4.951 GeV.

For the process of $e^+ e^- \to \Sigma^+ \bar{\Sigma}^-$ in the Born approximation, in Fig.~\ref{WideRange} we show the dependence of the
total cross as a function of the center-of-mass energy $\sqrt s$.
It is seen that at the energy of $\sqrt s$ = 3.4 GeV, the total cross section has a maximum value and then, with increasing energy,
the total cross section smoothly decreases.
A comparison with the experimental data shows that our result agrees well with the experimental points.
In Fig.~\ref{WideRange}, we constructed a plot according to the theoretical results for the total cross section in the Born approximation.
Our theoretical result is compared with the experimental data from the BESIII.

We would like to note that the propagator between the $D^0 \bar{D^0}$ vertices contains two types of hyperons -- $\Sigma^+_c, \Xi^+_c$,
that is, there are different options. These hyperons are obtained from different combinations of quarks.
When we performed the calculations, we took for the hyperon mass in the propagator the average value of the mass of the two hyperons.

It should be noted that the quantities $Z_D (s)$ from \eqref{ZD} and $Z_{3g}\br{s}$ from \eqref{Z3g} are considered the main parts of
the general cross section, which give the corresponding contributions ($D$-meson loop and three-gluons).
In Fig.~\ref{fig.ZD} we present the dependence of the quantity of $Z_D (s)$ on the total energy $\sqrt{s}$ in the range from the
reaction threshold $\sqrt{s}= 2M_\Sigma^+$ to $4.951~$GeV.
It can be seen from the figure that in the real and imaginary parts of $Z_D (s)$ the quantity remains
practically the same as in the case of the production of final states $p \bar {p}$ [Fig.~7(a) in \cite{Bystritskiy}],
and $\Lambda \bar {\Lambda}$ [Fig.~6 in \cite{BA}], and $\Sigma^0 \bar{\Sigma}^0$ [Fig.~6 in \cite{AAhmadov}].
Now in Fig.~\ref{fig.Z3g} we show the dependence of the real and imaginary parts of the quantity
$Z_{3g} (s)$ on the total energy $\sqrt{s}$ in the range from the reaction threshold
$\sqrt{s}= 2M_\Sigma^+$ to $4.951~$GeV.
We would like to note that in this work the general behavior of the curves of the quantity $Z_{3g} (s)$ by $\sqrt{s}$ is very different
from the curves, which were produced in the final state $p \bar {p}$ [Fig. ~7 (B) \cite{Bystritskiy}],
$\Lambda \bar {\Lambda}$ [Fig.~7 \cite{BA}] and $\Sigma^0 \bar{\Sigma}^0$ [Fig.~7 in \cite{AAhmadov}].
It can be seen from this graph that the characteristic large negative values of the quantity $Z_{3g} (s)$ remain the same,
which gives a large relative phase with respect to the Born contribution to the amplitude.
The resonance position $\psi(3770)$ in Figs. \ref{fig.ZD} and \ref{fig.Z3g} is marked by a vertical dashed line.
\begin{figure}
	\vspace{5mm}
	\centering
    \includegraphics[width=0.60\textwidth]{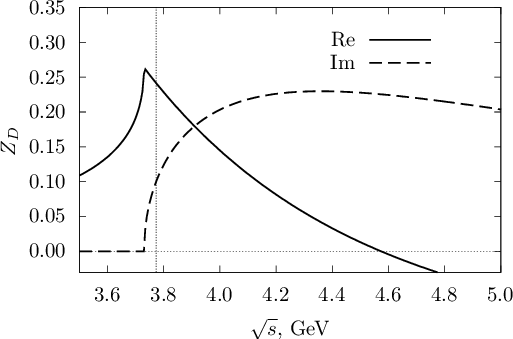}
    \caption{
    	The quantity $Z_D\br{s}$ from (\ref{ZD}) as a function of the total invariant energy $\sqrt{s}$ starting
        from the threshold $\sqrt{s} = 2M_\Sigma^+$.
        The position of $\psi(3770)$ is shown by a vertical dashed line.}
    \label{fig.ZD}
\end{figure}

\begin{figure}
	\vspace{5mm}
	\centering
    \includegraphics[width=0.60\textwidth]{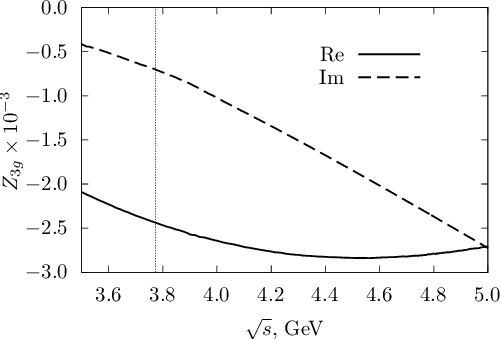}
    \caption{
    	The quantity $Z_{3g}\br{s}$ from (\ref{Z3g}) as a function of the total invariant energy $\sqrt{s}$ starting
        from the threshold $\sqrt{s} = 2M_\Sigma^+$.
        The position of $\psi(3770)$ is shown by a vertical dashed line.}
    \label{fig.Z3g}
\end{figure}
\begin{figure}
	\centering
    \includegraphics[width=0.55\textwidth]{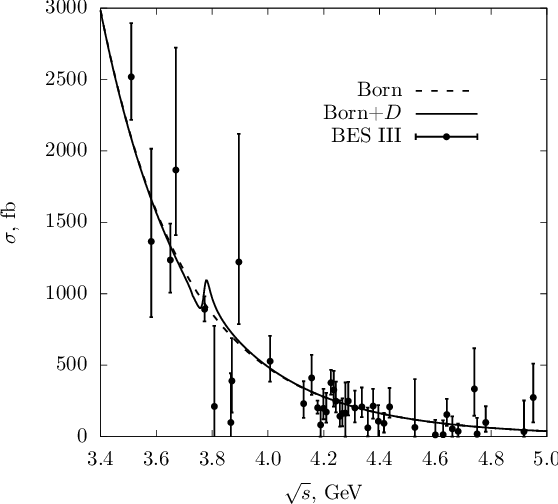}
    \caption{The total cross section as a function of the center-of-mass energy $\sqrt {s}$ with the inclusion of the
    $D$-meson loop contributing to the BESIII data \cite{I26}.}
    \label{fig.NarrowDLoop}
\end{figure}
\begin{figure}
	\centering
    \includegraphics[width=0.50\textwidth]{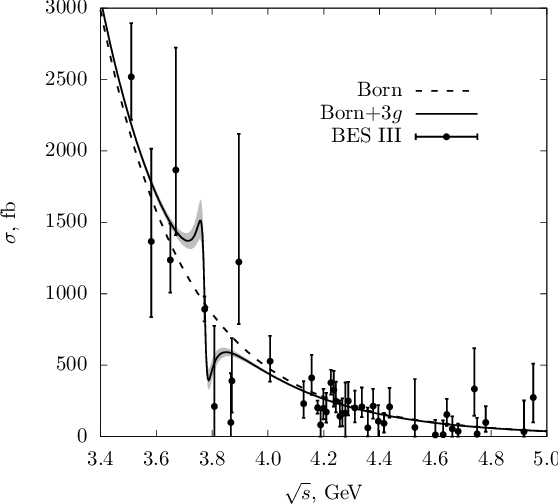}
    \caption{The total cross section as a function of the center-of-mass energy $\sqrt {s}$ with the inclusion of the
    three-gluons contributing to the BESIII data \cite{I26}.}
    \label{fig.Narrow3g}
\end{figure}
\begin{figure}
	\centering
    \includegraphics[width=0.50\textwidth]{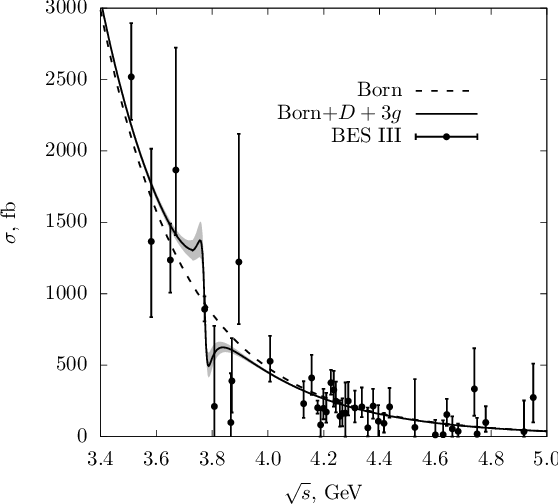}
    \caption{The total cross section as a function of the center-of-mass energy $\sqrt{s}$ including two mechanisms
        ($D$-meson loop and three gluon) in comparison with the BESIII data \cite{I26}.}
    \label{fig.CSNarrow2}
\end{figure}
The contributions of the pure $D$-meson loop to the Born cross section are shown in Fig.\ref{fig.NarrowDLoop}.
In Fig.\ref{fig.Narrow3g} we illustrate the result of a pure three-gluon contribution.
The theoretical results obtained for both of these contributions are compared with the data of the BESIII \cite{I26} collaboration.

It can be seen from Fig.~\ref{fig.NarrowDLoop} that the experimental result at the point $\sqrt {s} = 3770$ GeV, i.e.,
the $\psi(3770)$ charmonium resonance point, is very close to our theoretical result.
It is seen from Fig.~\ref{fig.Narrow3g} that the experimental result at the point $\sqrt {s} = 3770$ GeV, i.e.,
the resonance point of the charmonium $\psi (3770)$, completely coincides with our theoretical result.
It is seen from Fig.~\ref{fig.CSNarrow2} that the experimental result at the charmonium resonance
point $\psi(3770)$, which takes into account both contributions, i.e., the $D$-meson loop and the three-gluon mechanism,
completely coincides with our theoretical result.

In Fig.~\ref{fig.CSNarrow2} we present the results of the dependence of the total cross section in the energy range
$\sqrt{s}$ = 3.4 - 5.0 GeV taking into account the $D$-loop and three-gluon mechanisms in the Born cross section
and compare it with the BESIII \cite{I26} data.
In this graph, we want to show that the peak is near the resonance of the $\psi(3770)$ charmonia and is clearly visible.
It is necessary to note that in our model, we mainly consider the vicinity of $\psi(3770)$.
But, in addition to this, we consider the vicinity with new charmoniumlike states in the same energy region, for example,
$\psi(4040)$, $\psi(4160)$, $Y(4230)$, $Y(4360)$, $\psi(4415)$ and $Y(4660)$.
Thus, we basically have the $\psi(3770)$ charmonium in the intermediate state, but in addition we calculate the total
cross section taking into account the new charmoniumlike states in the intermediate state.

We would like to remind that in the process under consideration we do not make any additional parameter fitting.
We fix all the necessary parameters of our model by calculating the process $e^+e^- \to p\bar {p}$ \cite{Bystritskiy}.

Now in Fig.~\ref{fig.PhiNarrow} we illustrate the dependence of the total relative phase $\phi_\psi$ from the center of mass energy $\sqrt{s}$.
We determine the total relative phase $\phi_\psi$ as the contribution of the charmonium $\mathcal{M}_\psi$ to the amplitude with
respect to the Born contribution of $\mathcal{M}_B$ without taking into account the Breit-Wigner factor, that is,
\ba
    S_D (s) + S_{3g} (s) = |S\br{s}| e^{i \phi_\psi},
\ea
\begin{figure}
	\centering
    \includegraphics[width=0.70\textwidth]{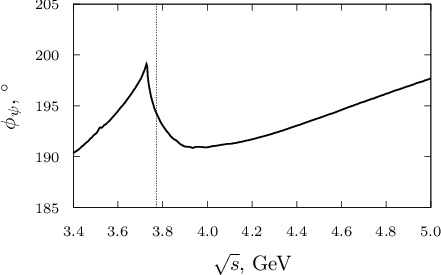}
    \caption{The relative phase of the charmonium $\psi(3770)$ contribution as a function of the total invariant energy $\sqrt{s}$.}
    \label{fig.PhiNarrow}
\end{figure}
where the quantities $S_D(s)$ was defined in \eqref{SDG} and $S_{3g}(s)$ is from \eqref{S3g}.
This figure shows that the resonance position $\psi(3770)$ is marked also by a vertical dashed line.
We would like to note that, as can be seen from Fig.~\ref{fig.PhiNarrow}, at the point of the $\psi(3770)$ charmonium
the relative phase and the corresponding total cross section \eqref{TotalCrossSection} are obtained
\ba
    \sigma_\psi = 1157.7~\mbox{fb},
    \qquad
    \phi_\psi = 194.4^{\circ}.
\ea
We can assume that such a property is common for the decay of charmonium into two baryons to the final state.
This was shown for the decay of the charmonium $\psi(3770)$ into a pair of $p\bar{p}$ and a pair of $\Lambda\bar{\Lambda}$
in the final states in \cite{Ahmadov,Bystritskiy,BA} and for the charmonium $\chi_{c2}(1P)(3556)$ in \cite{Kuraev}.

In this present work, in addition to the contribution of the $\psi(3770)$ charmonium in the intermediate state, we also considered
the contributions from charmoniumlike states, $\psi(4040)$, \,\,$\psi(4160)$, \,\,$Y(4230)$,\,\,$Y(4360)$,\,\,$\psi(4415)$,
and $Y(4660)$ in the $e^+e^- \to \Sigma^+\bar{\Sigma}^-$ process.
In Fig.~\ref{CharmoniumStates} we present the total cross section at the center-of-mass of energy in the region
$\sqrt{s}$ = 3.510 - 4.951 GeV according to the results of the contributions of charmonium (like) states,
$\psi(4040)$, \,\,$\psi(4160)$, \,\,$Y(4230)$,\,\,$ Y(4360)$,\,\,$\psi(4415)$, and $Y(4660)$ in the $e^+e^- \to \Sigma^+\bar{\Sigma}^-$
process, respectively.
We would like to note that in Fig.~\ref{CharmoniumStates} it is seen that the experimental result at the point for each resonance of charmonia
$\psi(4040)$, \,\,$\psi(4160)$, \,\,$Y(4230)$,\,\,$Y(4360)$,\,\,$\psi(4415)$, and $Y(4660)$ are in very good agreement with our theoretical result.
\begin{figure}
	\centering
    \subfigure[]{\includegraphics[width=0.38\textwidth]{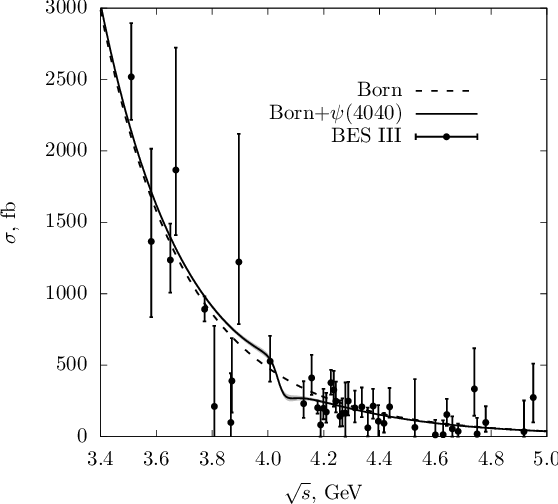}\label{psi4040}}
	\hspace{0.05\textwidth}
    \subfigure[]{\includegraphics[width=0.38\textwidth]{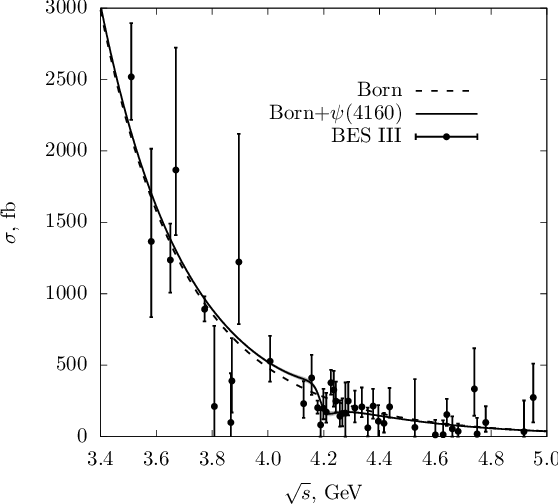}\label{psi4160}}
    \vspace{0.5cm}
    \centering
    \subfigure[]{\includegraphics[width=0.38\textwidth]{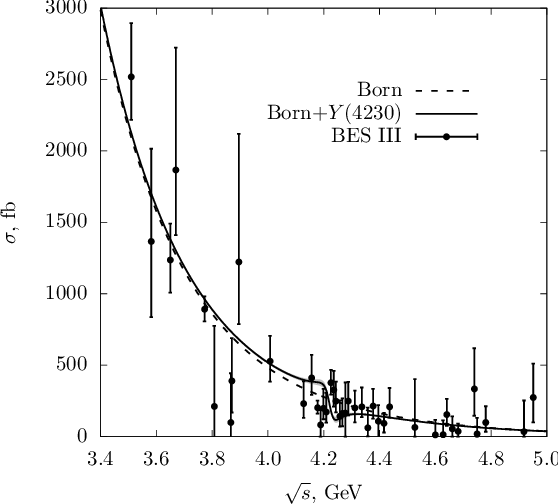}\label{Y4230}}
	\hspace{0.05\textwidth}
    \subfigure[]{\includegraphics[width=0.38\textwidth]{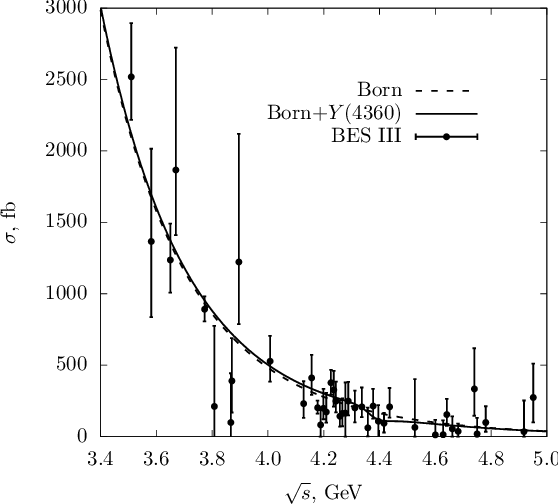}\label{Y4360}}
    \vspace{0.5cm}
    \centering
    \subfigure[]{\includegraphics[width=0.38\textwidth]{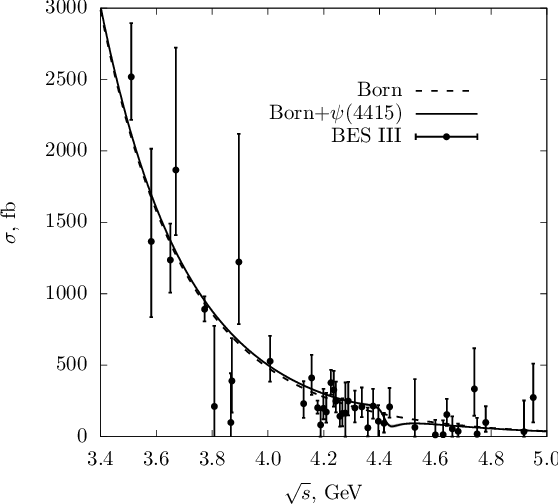}\label{psi4415}}
	\hspace{0.05\textwidth}
    \subfigure[]{\includegraphics[width=0.38\textwidth]{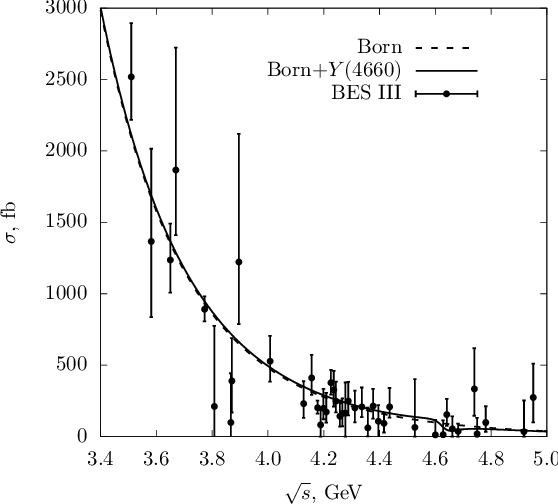}\label{Y4660}}
    \caption{The total cross section as a function of the center-of-mass energy $\sqrt{s}$ for the contribution of
    charmonium (-like) states, $\psi(4040)$, \,\,$\psi(4160)$, \,\,$Y(4230)$,\,\,$Y(4360)$,\,\,$\psi(4415)$, and $Y(4660)$,
    in comparison with the BESIII data \cite{I26}.}
    \label{CharmoniumStates}
\end{figure}

% =======================================================================
\section{Summary and Conclusion}
\label{sec.Conclusion}
% =======================================================================
In the present work, we have investigated the process of electron-positron annihilation into a $\Sigma^+ \bar{\Sigma}^-$
pair in the vicinity of the charmonium resonance $\psi(3770)$ for the energy of BESIII.
Moreover, we also studied this process in the vicinity of new charmoniumlike states in the
same energy region, such as $\psi(4040)$, $\psi(4160)$, $Y(4230)$, $Y(4360)$, $\psi(4415)$, and $Y(4660)$.
In the process $e^+ + e^- \to \Sigma^+ + \bar{\Sigma}^-$, besides the Born mechanism, which is presented in the framework
of QED, we also investigated two more contributions associated with the intermediate state of the charmonium $\psi(3770)$
and other new charmoniumlike states, such as $\psi(4040)$, $\psi(4160)$, $Y(4230)$, $Y(4360)$, $\psi(4415)$, and $Y(4660)$.
Thus, one of these contributions is a $D$ meson loop and the other is the contribution of a three-gluon mechanism.
Here we should remind that the process $e^+ + e^- \to\Sigma^+ + \bar{\Sigma}^-$ can also be initiated by the vector-charmonium
state as $\psi(3770)$, as well as other charmoniumlike states, which were mentioned above.
We understand that since the photon, $\psi$ and other charmoniumlike states are vector mesons, the structures of the
corresponding cross-section distributions is therefore similar.

It has been shown that in the present process both mechanisms make a significant contribution and account for
a large part of the final result.
It is also important that the curve we obtained reproduces the tendency of the experimental points in the left
and the right shoulders with respect to the central point.
We would like to point out once again that in this process we do not use any fitting procedures in our calculations.
All parameters were used fixed for the production of proton-antiproton channel in \cite{Bystritskiy}.
We wanted to be able to make an accurate scan with the small steps of the energy region around the charmonium $\psi(3770)$ resonance.
We can make a thorough conclusion that in the decay of charmonium, the phases of the
vertices $\psi \to p\bar{p}$, \, $\psi \to \Lambda\bar{\Lambda}$, \, $\psi \to \Sigma^0 \bar{\Sigma}^0$, and $\psi \to \Sigma^+ \bar{\Sigma}^-$
are large ($\phi_\psi \sim 200^\circ$) and can be accurately measured in these channels.
We would like to note that a large-phase generation was shown by us in the present work and in a series of
works \cite{Ahmadov,Kuraev,Bystritskiy,BA,AAhmadov}.
In conclusion, in the future we plan to consider other binary processes of formation of final states that are
induced by charmonium annihilation.

\section{Acknowledgements}
I am grateful to Dr. Yu. M. Bystritskiy for useful discussions.

\end{document}